\documentclass[aps,twocolumn,superscriptaddress,nofootinbib,prc,fleqn]{revtex4-2}

\usepackage{graphicx}
\usepackage{amsmath}
\usepackage{amssymb}
\usepackage{float}
\usepackage[colorlinks=true, pdfstartview=FitV, linkcolor=red,
citecolor=blue, urlcolor=blue]{hyperref}
\usepackage{graphicx}
\usepackage{dcolumn} 
\usepackage{bm} 
\usepackage{hyperref}
\usepackage{xcolor}
\usepackage{booktabs}

\newcommand{\PbPb}{\ensuremath{\mathrm{Pb+Pb}}}
\newcommand{\pp}{\ensuremath{\mathrm{pp}}}
\newcommand{\alphas}{\ensuremath{\alpha_\mathrm{s}}}
\newcommand{\Martini}{\textsc{martini}}
\newcommand{\Music}{\textsc{music}}
\newcommand{\IPG}{\textsc{ip-glasma}}
\newcommand{\Pythia}{\textsc{pythia}}

\newcommand{\RAA}{\ensuremath{R_\mathrm{AA}}}

\begin{document}
    \title{Leading order, next-to-leading order, and nonperturbative parton collision kernels: Effects in static and evolving media}
    \preprint{APS/123-QED}
    \author{Rouzbeh Modarresi Yazdi}
    \affiliation{Department of Physics, McGill University, 3600 University street, Montreal, QC, Canada H3A 2T8}
    \author{Shuzhe Shi}
    \affiliation{Center for Nuclear Theory, Department of Physics and Astronomy, Stony Brook University, Stony Brook, New York 11794–3800, USA}
    \affiliation{Department of Physics, McGill University, 3600 University street, Montreal, QC, Canada H3A 2T8}
    \author{Charles Gale}
    \affiliation{Department of Physics, McGill University, 3600 University street, Montreal, QC, Canada H3A 2T8}
    \author{Sangyong Jeon}
    \affiliation{Department of Physics, McGill University, 3600 University street, Montreal, QC, Canada H3A 2T8}

    \date{\today}
    
    \begin{abstract}
       Energetic partons traveling in a strongly interacting medium lose energy by emitting radiation and through collisions with  medium constituents. Nonperturbative, next-to-leading order and leading-order collision kernels are implemented within AMY-McGill formalism. The resulting gluon emission rates are then evaluated and compared by considering scattering occurring in a brick of quark-gluon plasma, as well as in a realistic simulation of \PbPb\, collisions at $\sqrt{s}=2.76$ ATeV using \Martini. We find that the variations in quenching of hard partons resulting from using different kernels can be important, depending on the overall value of the strong-coupling constant \alphas. 
    \end{abstract}

    \maketitle
    \section{Introduction}\label{sec:introduction}
The study of high-energy collisions of nuclei --- ``heavy-ions'' --- is the only practical way to study strongly interacting matter under extreme conditions of temperature and density, in terrestrial laboratories under controlled conditions. This active line of research seeks to understand quantum chromodynamics (QCD, the theory of the nuclear strong interaction) in all generality, and to answer fundamental questions about the phase diagram of QCD, and possibly to obtain precious information about the matter constituting the core of neutron stars, and the state of the universe a few microseconds after the Big Bang~\cite{Kapusta:2006pm}. 

A vibrant global research program is under way, and the flagship facilities at the experimental energy frontier are RHIC (the Relativistic Heavy-Ion Collider, at Brookhaven National Laboratory) and the LHC (the Large Hadron Collider, at CERN). Several experiments performed at those laboratories have now revealed an exotic state of matter: the quark-gluon plasma (QGP)~\cite{Jacak:2012dx}. In the QGP, the partons constituting nuclear matter are free to roam over distances commonly  considered macroscopic in hadronic physics. A considerable theoretical effort is currently being devoted to the detailed understanding of this QGP.  In this context, one of the theoretical breakthroughs of the last couple of decades has been the realization that the collective behavior of the QGP can be modeled very faithfully  by viscous relativistic fluid dynamics~\cite{[{See, for example, }][{, and references therein.}]Gale:2013da}. Hydrodynamics represents an effective theory of long wavelengths excitations, which can be extended to include nonequilibrium aspects with the introduction of transport coefficients such as shear ($\eta$) and bulk ($\zeta$) viscosities. The connection between the phenomenology of relativistic heavy-ion collisions, fluid dynamics, and QCD is provided by the equation of state which is evaluated nonperturbatively in lattice QCD~\cite{HotQCD:2014kol}. 

In addition to the measurements of one-body quantities like single-particle spectra of hadrons and of their collective motion --- through the extraction of flow coefficients \cite{Gale:2013da} --- there exists a class of {\em tomographic probes}. Those are penetrating and are able to reveal the inner workings of the QGP. Among those are QCD jets which form early in the heavy-ion collision, traverse and interact with the strongly interacting medium, fragment and are reconstructed by measuring their final states~\cite{Connors:2017ptx}. The very existence of QGP was revealed by  measurements that highlighted the jet-medium interaction, through the analyses of di-hadron correlations~\cite{Adams:2003,Nattrass:2016cln}. The measurements of jets, of how they lose energy to the medium, and the theoretical modeling of this rich dynamics is therefore a central theme in modern high-energy nuclear physics~\cite{Gyulassy:1993hr,Wang:1994fx,Baier:1996sk,Baier:1994bd,Baier:1996kr,Zakharov:1996fv,Wiedemann:2000za,Guo:2000nz,Wang:2001ifa,Zhang:2003yn,Schafer:2007xh,He:2015pra,Cao:2016gvr,Casalderrey-Solana:2014bpa,Shi:2018izg,Cao:2020wlm,Qin:2015srf,Majumder:2010qh,Armesto:2011ht,Qin:2009bk,Jeon:2003gi,Noronha-Hostler:2016eow,Andres:2016iys,Bianchi:2017wpt,Chien:2015vja,Andres:2019eus}.

The physics of the 
quark-gluon plasma is entering an era of precision science, where experiments and theory are precise and  mature enough to provide  quantitative characterization of this exotic  material. 
One the lessons learned over the past decade or so is that a realistic dynamical modeling  of the collisions system is key to the quantitative characterization of hot and dense strongly interacting matter. The current state-of-the-art approaches combine  partonic initial states, hydrodynamical evolution, and hadronic transport, while experimental final states are being connected with model parameters using techniques rooted in Bayesian statistics \cite{JETSCAPE:2020shq,JETSCAPE:2020mzn,Nijs:2020ors,Nijs:2020roc,Bernhard:2016tnd,Bass:2017zyn,Bernhard:2019bmu}. Those  elaborate hybrid approaches model the background against which the tomographic signals evolve, and whose properties they probe. Precisely because QGP studies are maturing, this work focuses on one specific element of this chain: it aims to focus on the finite temperature jet-medium interaction as calculated within the Arnold--Moore--Yaffe (AMY) framework, and to explore quantitatively the empirical consequences of a recent extension of its partonic scattering kernel beyond the perturbative treatment \cite{Moore:2021jwe,Schlichting:2021idr}. 

The plan of this paper is the following: in Sec.~\ref{sec:radiative_energy_loss} we discuss briefly the leading order radiative energy-loss rates and AMY-McGill formalism \cite{Jeon:2003gi,Turbide:2005fk} as implemented in \Martini\ \cite{Schenke:2009gb}, as well as the new next-to-leading (NLO) order and nonperturbative (NP) scattering kernels introduced above in the context of a ``brick test.'' In Sec.~\ref{sec:realistic_hydro} we embed the new rates in a realistic \Music\, hydrodynamic simulation of \PbPb\, collisions at $2.76$~ATeV. Finally in Sec.~\ref{sec:Conclusion} we discuss the results and their implications.

\section{Radiative Energy Loss and the Brick}\label{sec:radiative_energy_loss}
\subsection{Leading order rates}
    The Arnold--Moore--Yaffe (AMY) formalism was proposed to  calculate the gluon and photon emission from a thermalized plasma~\cite{Arnold:2001ba,Arnold:2001ms,Arnold:2002ja}. 
    The aim was to compute the full 
    leading order (in the strong coupling $g_s$) results for gluon (or photon) spectrum from a plasma, which meant the 
    inclusion of bremsstrahlung and inelastic pair-annihilation (or $1\rightarrow 2$ channels). 
    These channels are superficially at a higher order in \alphas\ than the Born contribution, but they receive parametric 
    enhancements from the collinear singularities which are regulated by effective thermal masses. 
    The end result of which is that these processes are found to be of the same order as the 
    $2\rightarrow 2$ channels. The inclusion of bremsstrahlung and inelastic pair-annihilation 
    then requires the treatment of the Landau--Pomeranchuk--Migdal (LPM) effect, which is done with dynamic (i.e., nonstatic) scatterers. 
    The complete leading order (LO) rates have been implemented in \Martini\,~\cite{Schenke:2009gb} 
    along with the energy-loss associated with elastic channels~\cite{Schenke:2009ik}. \Martini\, then solves the Fokker--Planck-type rate equations for the evolving distribution of energetic partons, and contain a gain and a loss term:
    \begin{align}
        \frac{dP\left(p\right)}{dt} = \int_{-\infty}^{\infty}\Big[P\left(p+k\right)&\frac{d\Gamma\left(p+k,k\right)}{dk} -\nonumber \\
        &P\left(p\right)\frac{d\Gamma\left(p,k\right)}{dk}\Big]\,dk,
        \label{eq:fokker_planck_master}
    \end{align}
    where $p$ is the momentum of the original energetic parton and $k$ is that of the emitted particle. 
    In this form, the LO scattering rates are given by~\cite{Schenke:2009gb,Qin:2008asm,Arnold:2008iy} 
    \begin{align}
        \frac{d\Gamma\left(p,k\right)}{dk} = &\frac{g^2_s}{16\pi\,p^7}\frac{1}{1\pm e^{-k/T}}\frac{1}{1\pm e^{-\left(p-k\right)/T}} \times \nonumber\\ 
        &\left\{\!\begin{array}{ll}
            C_F \frac{1+(1-z)^2}{z^3(1-z)^2}, & q\rightarrow qg\\
            \frac{1}{2}\frac{z^2+(1-z)^2}{z^2(1-z)^2}, & g\rightarrow q\overline{q} \\
            C_A\frac{1+z^4+(1-z)^4}{z^3(1-z)^3}, & g\rightarrow gg
        \end{array}\right\} \nonumber \times & \\
       & \int \frac{d^2\mathbf{h}}{\left(2\pi\right)^2}2\mathbf{h}\cdot{\rm Re}\,{\mathbf{F}\left(\mathbf{h},p,k\right)}\,,
       \label{eq:LO_AMY_RATES}
    \end{align}
    where $z\equiv k/p$ is the momentum fraction of the radiated particle, $\frac{1}{1\pm e^{-k/T}}$ and $\frac{1}{1\pm e^{-\left(p-k\right)/T}}$ are respectively the statistical factor for emitted parton and the remnant, which take into account the Bose enhancement ($-$) or Pauli blocking ($+$) effects, $\mathbf{h}\equiv(\mathbf{p}\times\mathbf{k})\times\widehat{\mathbf{p}}$ measures how collinear the emitted 
    particle is to the mother, and is taken to be parametrically of 
    $O\left(g_s T^2\right)$. The curly brackets contain the DGLAP parton splitting functions. Note that $\mathbf{F}\left(\mathbf{h},p,k\right)$ is the solution 
    to the following integral equation
    \begin{align}
        2\mathbf{h} = &i\,\delta E \left(\mathbf{h},p,k\right)\mathbf{F}\left(\mathbf{h}\right)+~ g^2_s\int\frac{d^2\mathbf{q}_{\perp}}{\left(2\pi\right)^2}C\left(\mathbf{q}_{\perp}\right)\times\nonumber\\
        &\Big\{\left(C_s-C_A/2\right)\left[\mathbf{F}\left(\mathbf{h}\right) -\mathbf{F}\left(\mathbf{h}-k~\mathbf{q}_{\perp}\right)\right] +\nonumber\\ &\left(C_A/2\right)\left[\mathbf{F}\left(\mathbf{h}\right)-\mathbf{F}\left(\mathbf{h}+p~\mathbf{q}_{\perp}\right)\right] + \nonumber\\
        &\left(C_A/2\right) \left[\mathbf{F}\left(\mathbf{h}\right)-\mathbf{F}\left(\mathbf{h}-(p-k)~\mathbf{q}_{\perp}\right)\right]\Big\},
        \label{eq:AMY_integral_equation_for_F}
    \end{align}
    where $\delta E$ is the energy difference between the initial and final states
    \begin{equation}
        \delta E\left(\mathbf{h},p,k\right) = \frac{\mathbf{h}^2}{2pk(p-k)} + \frac{m^2_k}{2k} + \frac{m^2_{p-k}}{2(p-k)} - \frac{m^2_p}{2p},
    \end{equation}
    and $m_i$ are the effective thermal masses induced by interactions with the medium. $C_s$ is the 
    quadratic Casimir operator which for quarks(antiquarks) and gluons takes the 
    values $C_s=C_F=4/3$ and $C_s=C_A=3$, respectively. In the case where 
    $g\rightarrow q+\bar{q}$, the prefactor $\left(C_s-C_A/2\right)$ should 
    multiply the term containing $\mathbf{F}\left(\mathbf{h}-p~\mathbf{q}_{\perp}\right)$.
    Finally, $C\left(\mathbf{q}_{\perp}\right)$ 
    is the transverse momentum broadening kernel: the differential rate of transverse 
    (to the hard parton) momentum exchange with the medium. Assuming that the medium is constituted of dynamical charges, its value at leading 
    order in \alphas\, is given by~\cite{Arnold:2008vd} 
    \begin{align}
        C_{\mathrm{LO}}\left(\mathbf{q}_{\perp}\right) &= \frac{g^2_s T^3}{q^2_{\perp}\left(q^2_{\perp} + m^2_D\right)}\int \frac{d^3p}{\left(2\pi\right)^3} \frac{p-p_z}{p}\times\nonumber\\
        &[2 C_A n_B(p)(1+n_B(p'))+\nonumber\\
        &4N_f T_f n_F(p)(1-n_F(p'))]\,,\nonumber\\ 
        m^2_D&=g^2_s T^2\left(\frac{N_c}{3}+\frac{N_f}{6}\right).
    \end{align}
    where $N_c$ is the number of colors in the theory and $N_f$ is the number of fermion flavors under consideration.
    $n_B$ and $n_F$ are respectively the Bose and Fermi distribution, while $p$ and $p' = p+\frac{\mathbf{q}_{\perp}^2+2\mathbf{q}_{\perp}\cdot\mathbf{p}}{2(p-p_z)}$ are respectively the momenta of the medium particle before and after the scattering.
    \begin{figure*}[!htbp]
        \centering
        \includegraphics[width=0.95\linewidth]{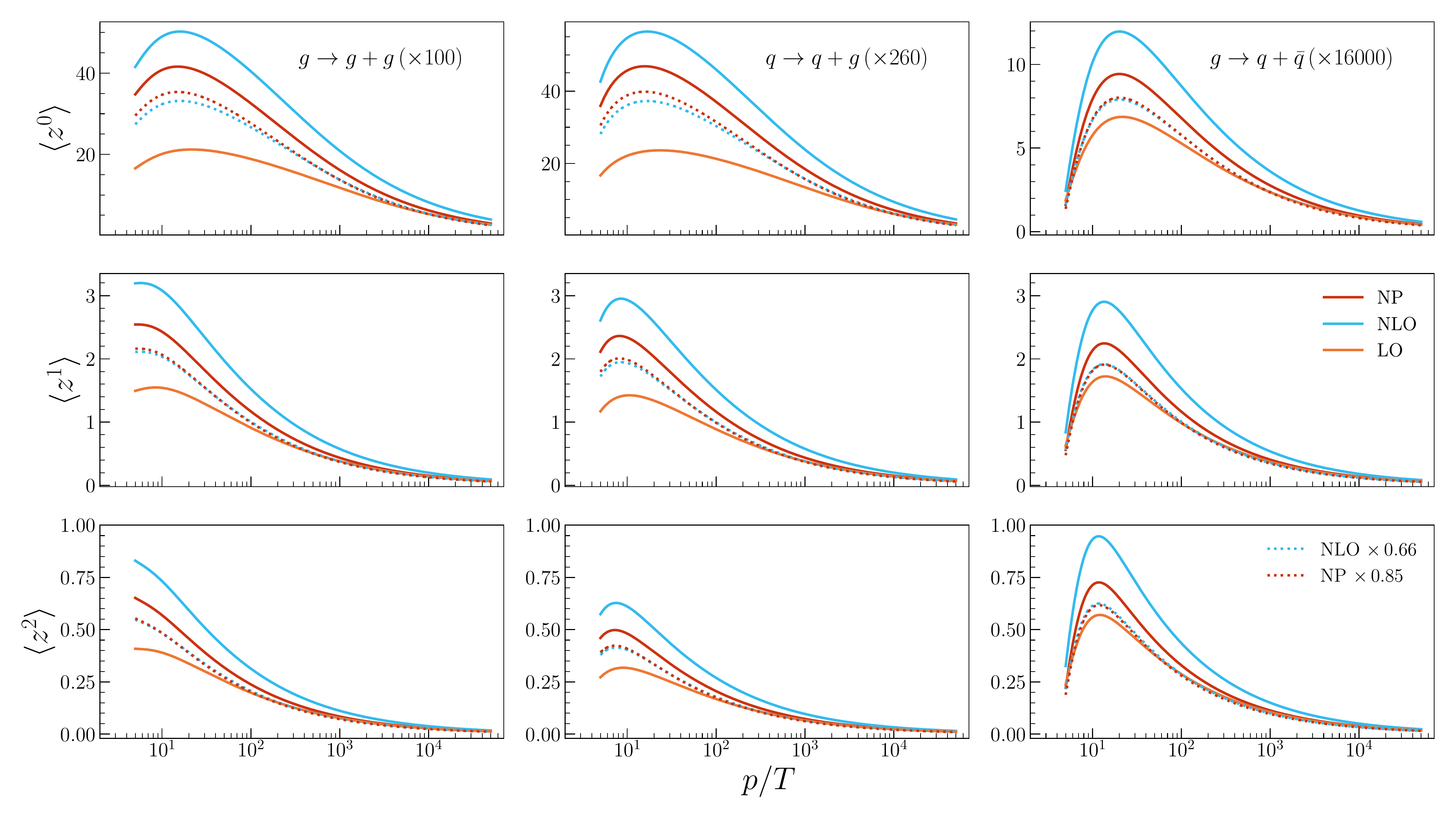}
        \caption{Momentum dependence of the zeroth, first and second moments of the integrated rate 
        (first, second and third row, respectively) for the three radiative processes plotted against temperature-scaled momentum of the incoming hard parton, $p/T$. The NLO and NP 
        dashed curves are multiplied by constant factors for a visual matching to the LO curve. Each column is multiplied by the number given in the top figure for clarity and in order to have a common $y$-axis for each row (except the first one).}
        \label{fig:fig1}
    \end{figure*}
\subsection{Jet-medium interactions at higher orders}
    More recently, efforts have been made to move beyond the LO calculation 
    of rates and collisional kernels. 
    The NLO scattering kernel has been computed 
    by Caron--Huot~\cite{Caron-Huot:2008zna} and used in a calculation of thermal photons~\cite{Ghiglieri:2013gia} as well as 
    for jet energy-loss in a medium within the AMY framework~\cite{Ghiglieri:2015ala}. Strongly coupled media, however,  may demand going beyond perturbation theory. 
    In what concerns numerical calculations on the QCD lattice it had been known that going to a dimensionally 
    reduced effective theory of QCD, known as \textit{EQCD} or 
    \textit{electrostatic QCD}~\cite{Braaten:1995ju}, enabled systematic higher-order calculations. Therefore, nonperturbative (NP) calculations of the elastic-scattering kernels 
    via lattice simulations of EQCD have recently been performed, allowing for NP QCD input to jet-medium 
    interactions and energ-loss. 
    The NP scattering kernel has been computed on the lattice~\cite{Moore:2019lgw} 
    for EQCD and matched to (3+1)D QCD for use in jet-medium energy-loss 
    calculations~\cite{Moore:2021jwe} for an infinite medium. Medium-induced splitting 
    rates were then computed using the new NP
    kernel and compared with to NLO and LO rates. Importantly, significant 
    differences were observed between the NP and LO results~\cite{Schlichting:2021idr}.
    
    In this work, we use the NLO and NP kernels developed in 
    Ref.~\cite{Moore:2021jwe} in Eq$.$~\eqref{eq:AMY_integral_equation_for_F} 
    and solve for the function $\mathbf{F}\left(\mathbf{h},p,k\right)$. 
    The results of those computations are then implemented in \Martini\, together with our own re-calculation of the LO rates \cite{Arnold:2002ja},  to quantify energy loss and to evaluate the consequences of the different scattering kernels. In the exploratory study performed here we use LO thermal masses, gluon emissions are still treated as collinear to the radiating parton, and collisional energy loss is treated at LO in the strong coupling \cite{Schenke:2009gb}.
    
    \begin{figure*}
        \centering
        \includegraphics[width=.9\linewidth]{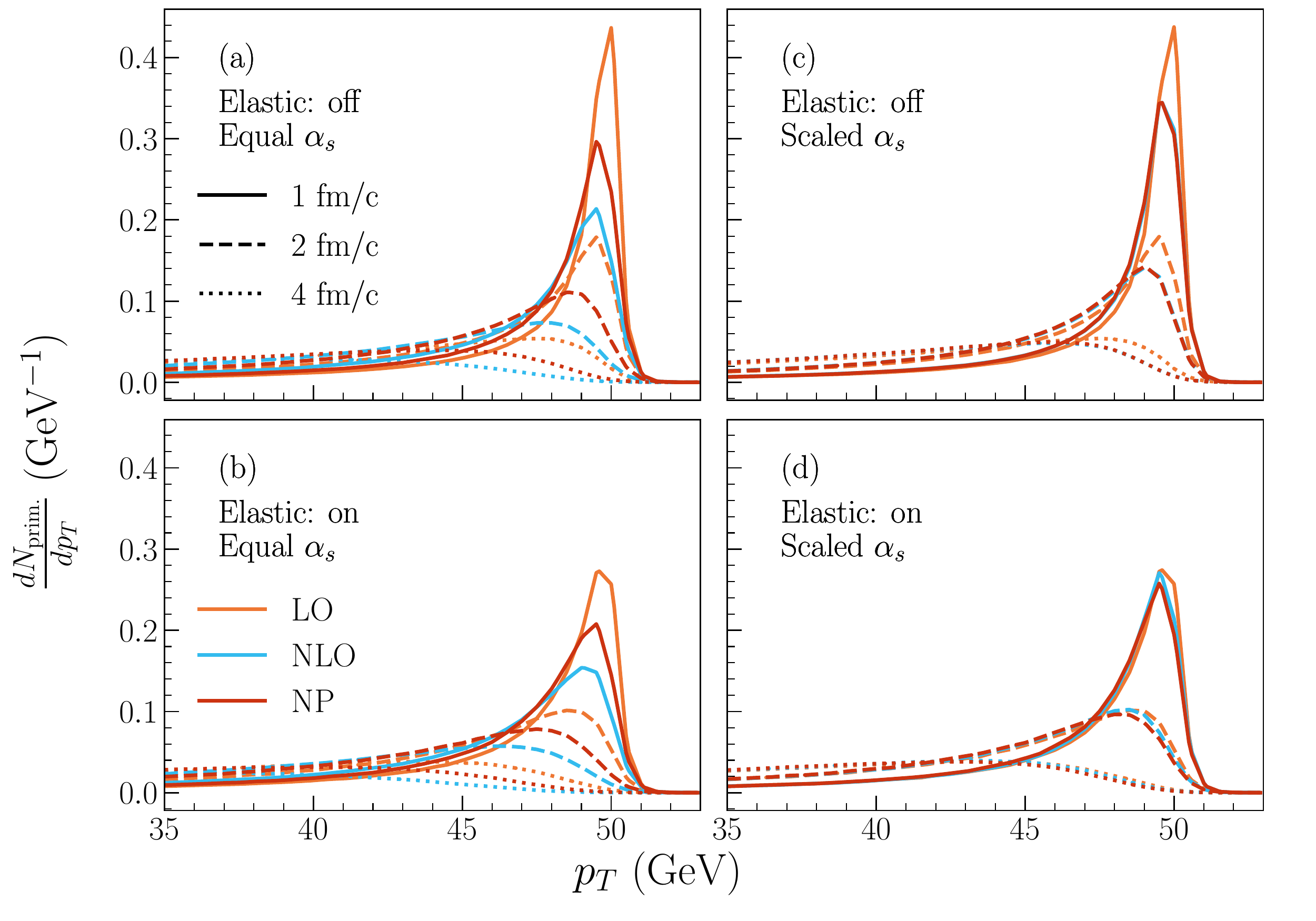}
        \caption{Evolution and broadening of the primary quark distribution computed by \Martini\, when initialized to $p=p_x=50$~GeV, traveling through a QGP brick of $T=0.3$ GeV. Elastic-scattering channels are turned off(on) in the upper(lower) panels. The left panels take equal coupling ($\alphas=0.3$) for all three rate sets, whereas the right panels use scaled values according to \protect{Eq.~\eqref{eq:scaled_coupling}}.}
        \label{fig:fig2}
    \end{figure*}
    \begin{figure}[htp]
        \centering
        \includegraphics[width=\linewidth]{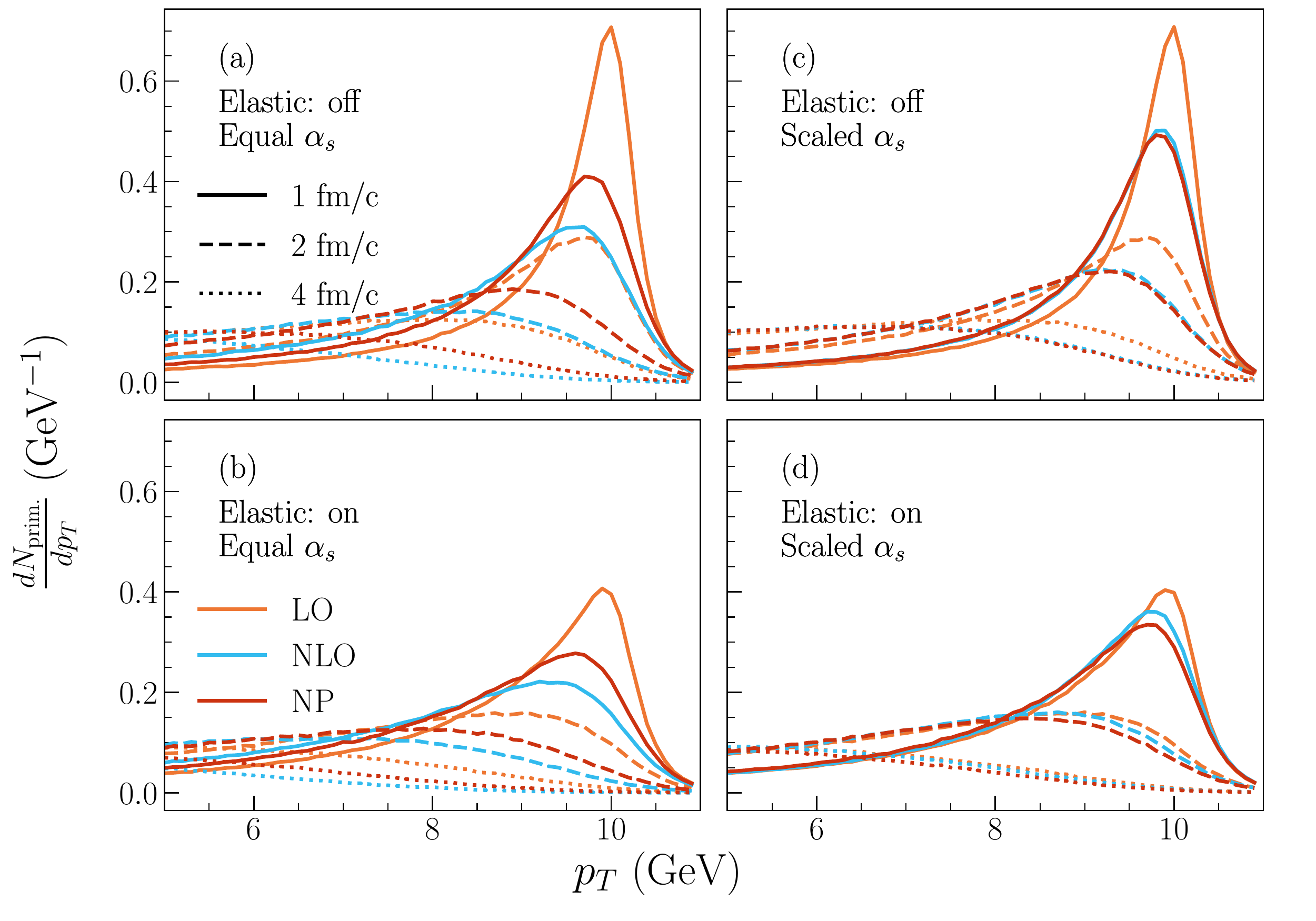}
        \caption{Same as \protect{Fig.~\ref{fig:fig2}} but for quarks with initial momentum $p=p_x=10$~GeV.}
        \label{fig:fig3}
    \end{figure}
    
\subsection{Brick test of rates with LO, NLO and NP Collision Kernels }\label{subsec:comp_rates_and_brick_test}
    A first step in comparing the different scattering kernels is assessing the energy-loss rate. As is known from elementary statistics, very useful general information about a given distribution function may be obtained by studying the behavior of its moments. In this spirit, we consider the integrated rates and 
    some of its  moments
    \begin{align}
        \langle z^i \rangle \equiv \frac{1}{g^4_s T} 
        \bigg( &\int_{0^{+}}^{1} |z|^i \frac{d\Gamma}{dz} dz -%\nonumber\\
        \int_{-\infty}^{0^{-}} |z|^i \frac{d\Gamma}{dz} dz \bigg),
        \label{eq:moments_of_rates}
    \end{align}
    where $z\equiv k/p$ is the ratio between the momentum of the emitted or absorbed particle and that of the original energetic parton\footnote{The emission and absorption occur when $z>0$ or $z<0$, respectively.}. 
 Then, $\langle z^0 \rangle g_s^4T\Delta t$ and $\langle z^1 \rangle g_s^4T\Delta t$ are respectively the difference between the number of parton emissions and that of parton absorptions, and the net energy-loss fraction divided by incident energy (``energy-loss fraction'' from here on)
 within time interval $\Delta t$. Also,  $\frac{\langle z^2 \rangle}{\langle z^0 \rangle} - \big(\frac{\langle z^1 \rangle}{\langle z^0 \rangle}\big)^2$ gives the variance of energy loss ratio per event, and  $\Gamma$ is the rate, computed using the LO, NLO or NP collision kernel. In what follows we refer to rates calculated using the NLO and NP kernels embedded in Eq.~\eqref{eq:LO_AMY_RATES} as NLO or NP rates.

    Figure~\ref{fig:fig1} compares the moments of the rates for 
    $q\rightarrow q+g$, $g\rightarrow g+g$ and $g\rightarrow q+\bar{q}$ 
    with the strong coupling $g_s$ and a factor of temperature scaled out for the 
      LO, NLO and NP scattering kernels. At a given value of the scaled momentum, the magnitude of the net scattering rate evaluated with the  NP kernel sits between that obtained with the LO and NLO expressions, clearly suggesting a slow convergence of the perturbative expansion. The slow convergence of many quantities evaluated in the framework of finite-temperature QCD is a common occurrence \cite{Caron-Huot:2008zna,Kapusta:2006pm, Ghiglieri:2013gia}.  
      Depending on the value of the specific momentum (momentum divided by temperature) and on the moment, the results obtained with the LO, NLO, and NP kernels can differ by as much as 100\%, as seen in Fig.~\ref{fig:fig1}
      
    Observing the different moment distributions, the similarity of shapes obtained with the different kernels suggests a rescaling procedure. We therefore re-scale the curves by a constant factor, adjusted to match the ultraviolet behavior of the LO rate, and obtain the results shown as dashed curves in Fig.~\ref{fig:fig1}. 
    It can be observed that the resulting scaled NP and NLO curves are very close to each other 
    and match neatly with the high-momentum tail of the LO order rate. 
    This is to be 
    expected given the construction of NLO and NP kernels~\cite{Ghiglieri:2015ala,Moore:2019lgw}. 
    The major difference between the scaled rates and the LO result is for $p\approx 10\,T$, where NLO and NP results can now be $10\%$ (gluon to quark splitting) to $40\%$ (gluon emission from gluon) higher than the LO rate, judging by the $\langle z^1\rangle$ panels (i.e., the net energy-loss fraction).
    One can interpret this rescaling of the rates as using different values of \alphas, scaled appropriately relative to each other for the different rate sets. In fact, it is a common practice in phenomenological studies to tune \alphas\, according some observable that measures the medium modification of energetic partons such as charge hadron modification factor ($R_{\rm AA}^{h^\pm}$). Importantly, the fact that NLO and NP rates predict more quenching than the LO one at low transverse momentum is of phenomenological interest, and therefore of consequence.
    It is known~\cite{Park:2021yck} that simulations taking LO AMY rates with fixed \alphas\, predicted a $p_T$-slope of $R_{\rm AA}^{h^\pm}$ that is more flat  
    than the experimental results. 
    Calculations with NLO and NP scattering kernels enhance the energy-loss for low momentum partons, which open the possibility of better agreement with experimental data.
    
    To see the differences in a more detailed way, we 
    perform a ``brick test,'' using \Martini\, with the three rate sets. The brick is an idealized static QGP medium of infinite extent at a constant temperature. A quark propagates in the brick, and one studies how the medium influences the parton energy. Realistic nuclear collisions often create finite media with density and flow gradients: the brick dispenses with those complications in order to establish a baseline scenario.  
    In what follows, unless stated otherwise, the traveling parton is a quark. We start 
    with a parton of $\mathbf{p}=\left(50,0,0\right)\mathrm{GeV}$ traveling within a 
    QGP brick of temperature $T=0.3$~GeV taking snapshots of its evolving spectrum. This representation of the time evolution enables an appreciation of how quickly the medium can transform the original $\delta$ function at a transverse momentum of 50 GeV. 
    Figure~\ref{fig:fig2}(a) demonstrates the comparison 
    of the rate sets when evaluated at the same value of strong coupling, $\alpha_s = 0.3$, a value appropriate for phenomenology at the BNL Relativistic Heavy Ion Collider (RHIC) and at the CERN Large HAdron Collider (LHC). 
    The pattern of Fig$.$~\ref{fig:fig1} where the 
    NLO rate has a higher radiation rate relative to LO and NP is visible here, 
    particularly for longer travel times where the quark jet subjected to NLO 
    rates is clearly melting faster in the plasma. 
    Figure~\ref{fig:fig2}~(b) implements the same 
    conditions as Fig$.$~\ref{fig:fig2}~(a) but now 
    allowing for elastic scattering. The inclusion of collisional energy-loss 
    brings the distributions closer to each other without altering the observed hierarchy of energy-loss shown in Fig$.$~\ref{fig:fig2}~(a).
    
    Scaling the value of \alphas\, for the brick test, as done before with the integrated 
    rates in Fig$.$~\ref{fig:fig1} according to 
    \begin{equation}
        \alpha^{\text{NLO}}_s = \sqrt{0.66}\times \alpha^{\text{LO}}_s,~~ \alpha^{\text{NP}}_s = \sqrt{0.85}\times \alpha^{\text{LO}}_s,
    \label{eq:scaled_coupling}
    \end{equation}
    makes the NP and NLO curves nearly indistinguishable from each other in Fig$.$~\ref{fig:fig2}~(c). 
    After the scaling, the LO curve is well separated from the other two, particularly around the original jet momentum.
    Inclusion of the elastic-scattering channels for the case where the 
    NLO and NP rates use a scaled value for \alphas, as can be seen in Fig$.$~\ref{fig:fig2}~(d), 
    removes most of the difference between the LO rate and NLO, NP rates. There are 
    some differences that remain for longer time spectra, in particular 
    the NLO and NP curves slightly separating from each other while LO curve comes 
    closer to both. While this was expected from Fig$.$~\ref{fig:fig1} where the 
    largest difference between the scaled (NLO, NP) rates and the LO result is 
    expected to be for the smaller values of $p/T$, the rates plotted in that 
    figure still imply some difference for $p\approx 100\,T$, 
    even with a scaled 
    \alphas. We can push the parton jet energy to the mini-jet domain while keeping 
    the temperature constant and thus probe the region of the rates where the 
    differences are more pronounced. 
    Figure~\ref{fig:fig3} is the equivalent calculation for a $10$~GeV parton 
    travelling through the same medium. Unlike the more energetic parton 
    before, the difference between the rates is more clearly visible and  
    they persist even after
    the inclusion of collisional energy-loss and rescaling of the coupling. This further reinforces the observation that the major difference between the rates lies in the infrared region.
    
    We can further study the effect of the changing of the rates on the energy-loss of a energetic parton through a brick by 
    considering the fractional energy-loss
    \begin{equation}
        \frac{p_i - \langle p_T \rangle}{p_i} = \frac{1}{p_i} \left(p_i - \int p_T \frac{dN}{dp_T} dp_T\right),
        \label{eq:FEL_observable}
    \end{equation}
    where $p_i$ is the initial momentum of the primary parton.
    \begin{figure*}[!htp]
        \centering
        \includegraphics[width=0.8\linewidth]{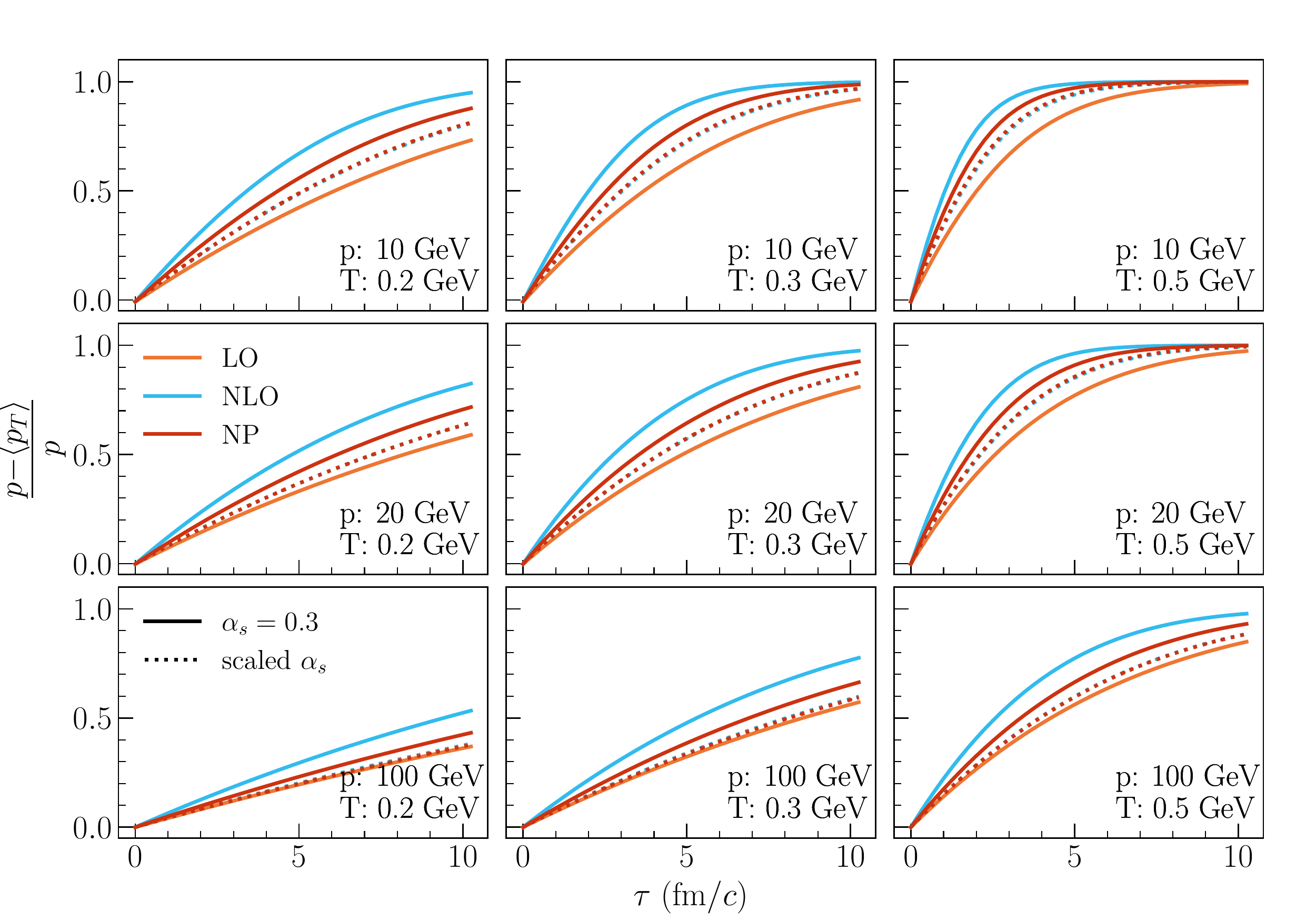}
        \caption{Fractional energy-loss of parton jets of momenta $p=\left(10,20,100\right)$ GeV (top to bottom) computed at three 
        temperatures of the QGP brick $T=\left(0.2,0.3,0.5\right)$ GeV (left to right). No collisional energy-loss is allowed. 
        Solid lines are $\alpha_s=0.3$ curves while the dotted lines denote the scaled values using \protect{Eq$.$~\eqref{eq:scaled_coupling}}.
        }
        \label{fig:fig4}
    \end{figure*}
    Figure~\Ref{fig:fig4} provides a comparison of the 
    un-scaled ($\alpha_s=0.3$) and scaled NLO and NP curves to the LO result 
    using the fractional energy-loss as defined in Eq$.$~\eqref{eq:FEL_observable}. 
    Without collisional energy-loss we can clearly see that as the energy of 
    the primary parton is reduced or temperature is increased the NLO and 
    NP curves separate more from the LO result even when scaled. The closest 
    the (NLO, NP) results get to the LO curve is for $p=100$~GeV and $T=0.2$~GeV ($p/T=500$) 
    which is in agreement with Fig$.$~\ref{fig:fig1}. 
    \begin{figure*}[!htp]
        \centering
        \includegraphics[width=0.8\linewidth]{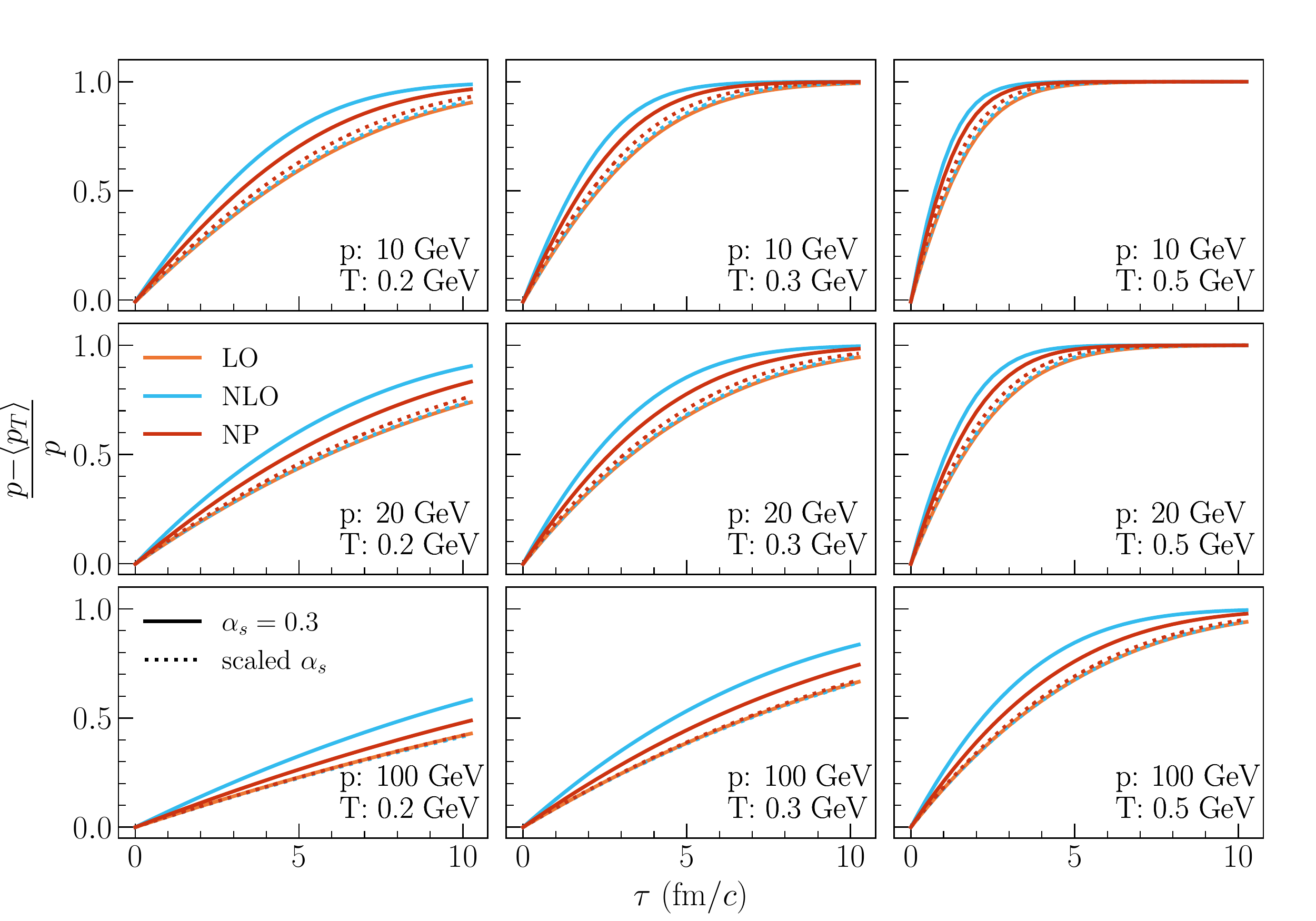}
        \caption{Same as  Fig$.$~\ref{fig:fig4} but with elastic channels turned on.}
        \label{fig:fig5}
    \end{figure*}
    Allowing for elastic scatterings as well as radiation in Fig$.$~\ref{fig:fig5}, 
    while introducing a small difference between the NLO and NP curves, also reduces the difference between the new rates and the LO result. 
    We can conclude that while for a 
    fixed value of \alphas\, the NLO curve predicts more quenching relative to 
    NP and LO rates, after scaling \alphas\, for each rate set we observe slightly more 
    quenching for the NP kernel relative to others at lower values of $p$. This is 
    marginally more pronounced when elastic scatterings are turned on. 
    \section{Energy loss and realistic evolution}\label{sec:realistic_hydro}
        We now employ and study the new rates in a calculation of charged hadron nuclear modification factor,
        \begin{align}
            R_{AA}^{h^\pm}(p_T) \equiv \frac{d\sigma_{AA}^{h^\pm}/dp_T} {N_\mathrm{bin}\,d\sigma_{pp}^{h^\pm}/dp_T}\,,
        \end{align}
        in an event-by-event simulation using realistic hydrodynamic backgrounds.
        $N_\mathrm{bin}$ is the number of binary collisions in a nucleus-nucleus collision.
        We focus on $0-5\%$ most central \PbPb\, collisions at beam energy $2.76$~ATeV and compare with the corresponding experimental data.

        The hydrodynamic background used in this study is provided by 
        \Music\,~\cite{Schenke:2010nt}, a $\left(3+1\right)$D relativistic viscous 
        hydrodynamic simulator of heavy-ion collisions. \Music\, concurrently solves 
        the conservation equation of energy-momentum tensor and the relaxation 
        equations of shear stress tensor and bulk viscous pressure~\cite{Molnar:2013lta,Denicol:2014vaa}.
        In this work, the fluid dynamics initial condition, at $\tau=0.4~\mathrm{fm}$, of the 
        stress-tensor is provided by two-dimensional \IPG\,calculation~\cite{Schenke:2012hg,Schenke:2012wb}, 
        and then the hydrodynamic quantities are evolved taking the HotQCD Collaboration equation of 
        state~\cite{HotQCD:2014kol}, a constant shear viscosity to entropy density $\eta/s=0.13$, and 
        the same parametrization for the bulk viscosity as in Ref.~\cite{Ryu:2015vwa}.  \newline \\

        The medium is then calculated in an event-by-event basis using \Music\, and 
        stored as input for \Martini. Hard partons are evolved by \Martini\, 
        through the medium and can lose (or gain) energy via their interactions 
        with it. In this work we do not account for the medium response to the 
        passage of energetic particles. The initial hard parton distribution is 
        determined by \Pythia~$8.209$~\cite{Sjostrand:2014zea} using the CMS 
        underlying event tune (tune \# 15)~\cite{CMS:2015wcf}. For our baseline \pp\, 
        calculations we use the \textsc{CTEQ 6L1}~\cite{Stump:2003yu} proton PDF while 
        the \PbPb\, results are computed using \textsc{nCTEQ15-np}~\cite{Kovarik:2015cma} 
        nuclear PDFs. All PDF sets are provided by the \textsc{LHAPDF}~$6.1.6$
        ~\cite{Buckley:2014ana} package. While the hard scatterings are generated by 
        \Pythia, their location is sampled in the transverse plane using the initial 
        energy density as provided by \IPG. \Pythia\, then performs a full vacuum 
        shower and passes the particle list to \Martini\, for evolution in the medium. 
        It should be noted that the partons free-stream until the QGP medium is 
        ``formed'' around them at $\tau=\tau_0$ which for us is taken to be 
        $\tau_0=0.4\,\mathrm{fm}/c$, the start time of the hydro.
        Evolution of momentum for a given parton is stopped if it can not be distinguished from a medium particle -- defined here by the condition $p<4\,T$ -- or it is not in the fireball. Finally, energy-loss of jets in the hadron-cascade stage is not taken into account in this work. While preliminary studies of energy-loss in this stage indicate potentially significant contribution for hadron transverse momenta between $2-10$ GeV\cite{Dorau:2019ozd} (outside of the $p_T$ range considered here), the interaction rates are much lower due to lower temperatures and therefore subdominant to energy-loss in the QGP stage.

        For the sake of keeping this study simple, there are features we do not include in this work and leave for future consideration: correlated background subtraction (recoil effects) and coherence (formation time). Recoil effects concern the inclusion of medium particles that receive energy from energetic partons in an 
        elastic-scattering event and get ``promoted'' to a jet particle. These, then, leave a hole in the medium which has to be subtracted at analysis level from the hadron and jet spectra. The main contribution of including recoil effects is in the better description of the jet transverse profile, an observable which we do not consider here. Their effect on the charged hadron \RAA\, is restricted to the low $p_T$ region of the spectrum, below what we study here\cite{Park:2021yck}. 
        Formation time effect is time taken by the medium to distinguish the radiated parton from the incoming jet. This effect shows itself in the time dependence of radiative rates \cite{Caron-Huot:2010qjx}. Finally, the different \alphas\ appearing in this study do not run. After all partons are frozen out, they are hadronized using the Lund string model as provided by \Pythia.

        \begin{figure}[htb]
            %\centering
            \begin{center}
                \includegraphics[width=0.9\linewidth]{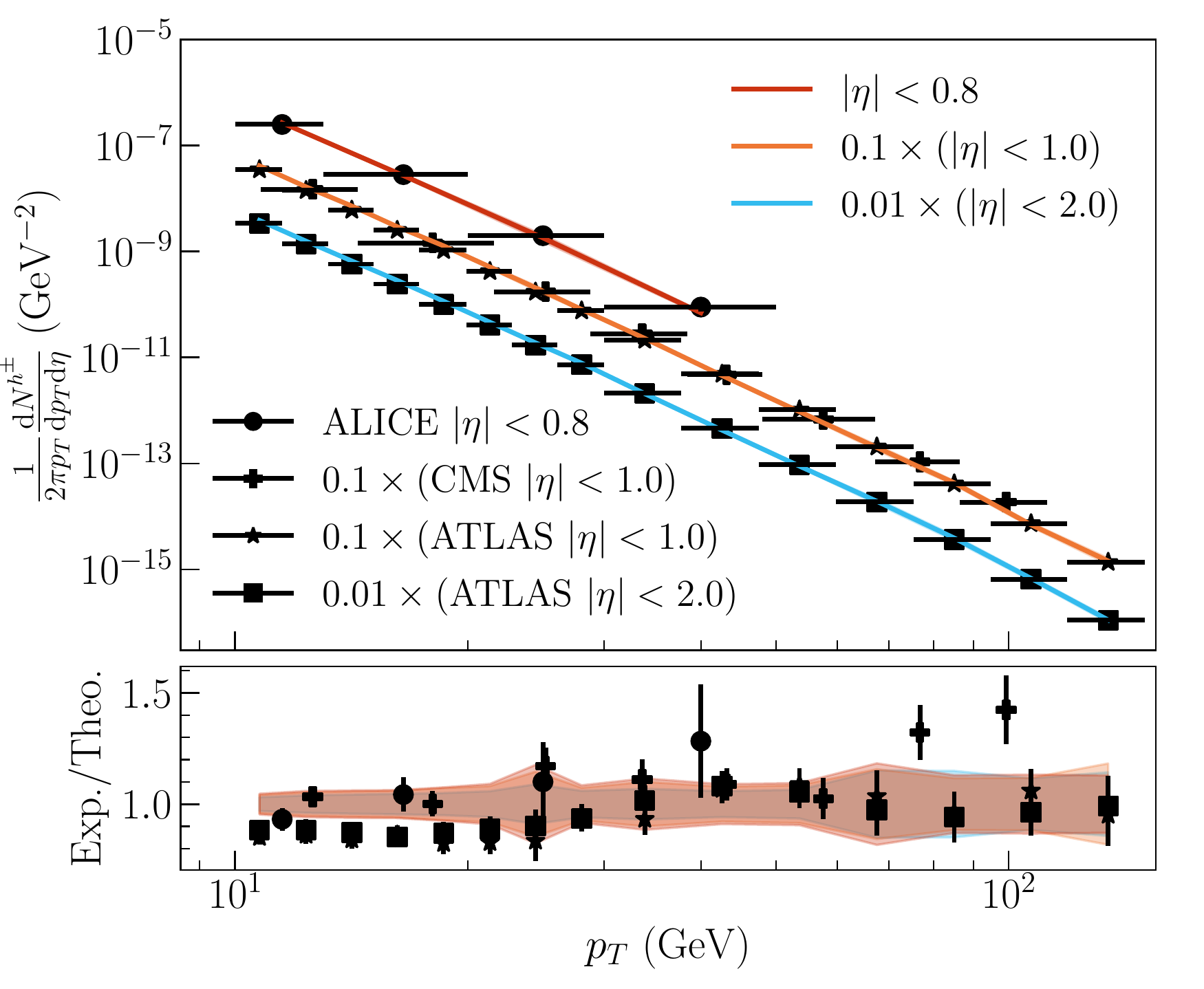}
                \caption{(upper panel) Invariant charged hadron yield computed using \Martini\, in proton-proton mode, 
                compared with data in three different pseudorapidity windows from ATLAS~\cite{ATLAS:2015qmb}, 
                CMS~\cite{CMS:2012aa}, and ALICE~\cite{ALICE:2018vuu} Collaborations. (lower panel) The ratio of data over theory is plotted on a linear scale. The shaded bands reflect the statistical uncertainties in the theoretical results.}
                \label{fig:fig6}
            \end{center}
        \end{figure}

        We present a sample of our baseline proton-proton calculation in 
        Fig$.$~\ref{fig:fig6}. As can be seen, we have good agreement with the experimental measurements 
        in three pseudorapidity windows, with data falling within $25\%$ of the 
        computed curves (lower panel).

        \begin{figure*}
            \centering
            \includegraphics[width=0.75\linewidth]{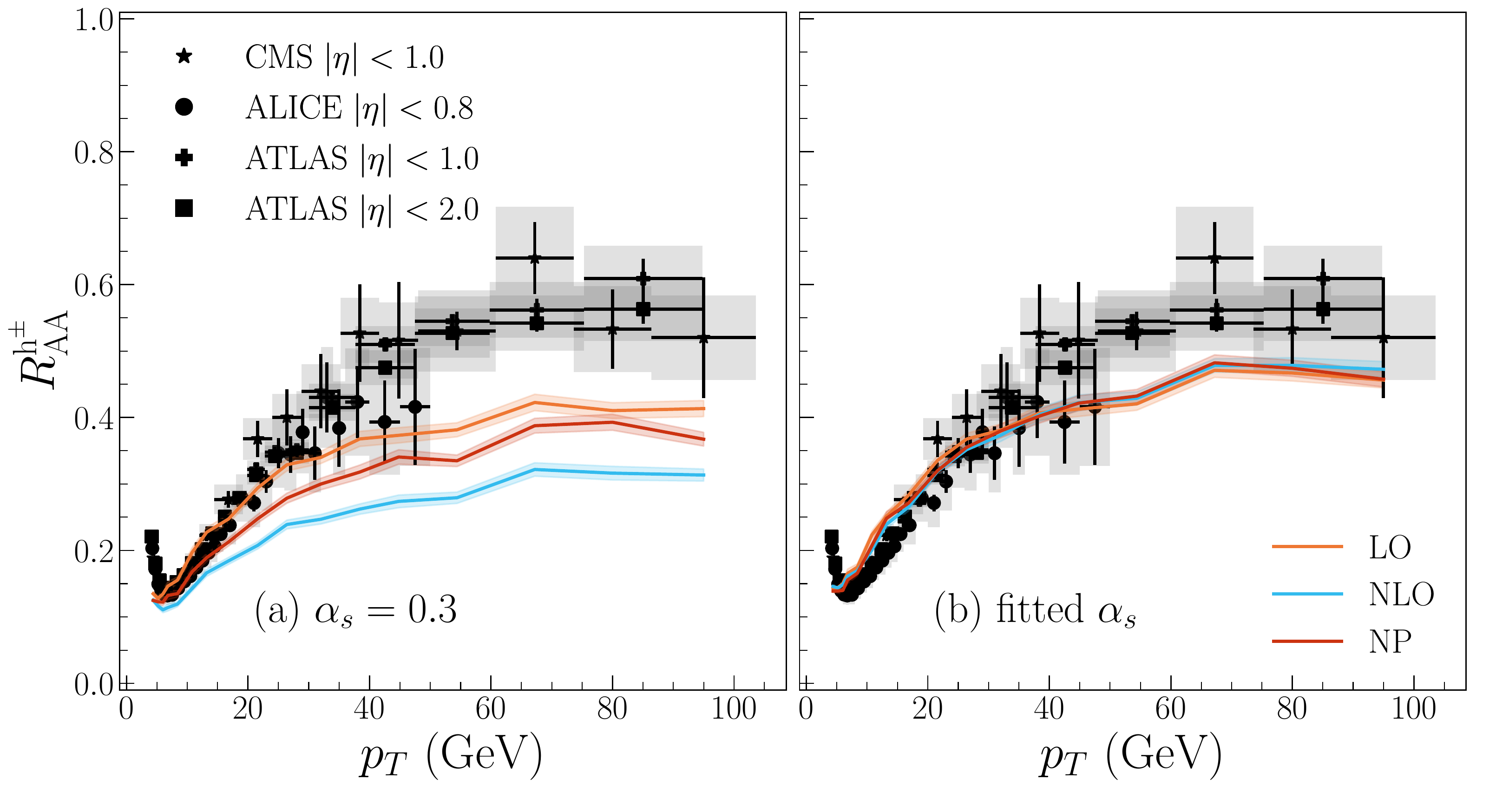}
            \caption{Comparison of the calculated charged hadron $R_{AA}$ to three different experimental results at midrapidity. The curves correspond to \Martini\, with the specified rate set for radiative energy loss. On the left, panel (a) uses $\alpha_s=0.3$ for all three rate sets while panel (b) on the right uses fitted values for \alphas. Experimental data are from ATLAS~\cite{ATLAS:2015qmb}, CMS~\cite{CMS:2012aa} and ALICE~\cite{ALICE:2018vuu}. The shaded bands reflect the statistical uncertainty in the theoretical results.}
            \label{fig:fig7}
        \end{figure*}
        Figure~\ref{fig:fig7}~(a) shows the result of a \Martini\, calculation using a constant 
        strong-coupling $(\alphas=0.3)$, for the three rate sets and including elastic energy loss. The pattern observed during the brick tests 
        also continues here. The LO and NLO rates induce the least and the most quenching, respectively, with the NLO curve being $\approx 75\%$ of the 
        LO result (averaged over the $p_T$ range under consideration). NP curve stands at $\approx 88\%$ of the LO result.

        We then tune \alphas\, separately for all three rate sets, in order to fit the experimental data of $R_{AA}$ by minimizing the chi-squared function,
        \begin{equation}
            \chi^2/\text{d.o.f} = \sum_i \frac{\left(y_{\text{expt,i}} - y_{\text{theor,i}}\right)^2}{\sum_s \left(\sigma_{\text{s,i}}\right)^2}/\sum_i 1,
            \label{eq:chi_squared_eq}
        \end{equation}
        where the sum over $i$ is over the experimental data points for $p_T > 10$ GeV and the sum over $s$ adds up all uncertainties. 
        \begin{table}
            \centering
            \begin{tabular}{lccl}\toprule
            Rate Set & $\alpha_s$ \\ \midrule
            LO & 0.280\\
            NLO & 0.242\\
            NP & 0.260 \\\bottomrule
            \end{tabular}
            \caption{Fitted values of \alphas\, from the $\chi^2$ fit.}
            \label{tab:fitted_alphas_table}
        \end{table}
        For the fitting procedure we use the experimental $R_{AA}$ in most central $0-5\%$ $2.76$~TeV \PbPb\, collisions from 
        ALICE~\cite{ALICE:2018vuu}, ATLAS~\cite{ATLAS:2015qmb} and CMS~\cite{CMS:2012aa}. With the fit results listed in Table$.$~\ref{tab:fitted_alphas_table}, we note that the ratios 
        between the fitted values of the strong-coupling are quite similar to the ratios predicted in Eq$.$~\eqref{eq:scaled_coupling},
        \begin{align}
            \sqrt{0.66} \lesssim &\frac{\alpha^{\text{NLO}}_s}{\alpha^{\text{LO}}_s} = 0.86\,,\nonumber\\
            \sqrt{0.85} \lesssim &\frac{\alpha^{\text{NP}}_s}{\alpha^{\text{LO}}_s} = 0.93\,.
        \end{align}
        It confirms the understanding that difference between the three rate sets can be mostly 
        resolved by rescaling the coupling constants. Meanwhile, the fact that they are slightly 
        above the predicted values reveals the effect of elastic scattering, which is 
        treated in the same way for all three rate sets.

        Fig$.$~\ref{fig:fig7}~(b) presents the results using fitted values for \alphas\, in each rate set. 
        With a fitted \alphas\, the NLO and NP curves give nearly identical results to the LO results. The expected enhancement of $p_T$-slope was not observed except the slight, although systematic, difference for $p_T<30$~GeV. There are two reasons for this. First, as shown in Fig$.$~\ref{fig:fig1}, one would expect a difference for partons with momentum $p\lesssim 30\,T \sim 10$~GeV. The fragmented hadron from this parton is not in the region of interest in phenomenological studies. Second, the difference shown in Fig$.$~\ref{fig:fig1} is the case only if there are no elastic scatterings present. In the presence of elastic scatterings, we could already see that the three rate sets would give very similar results (see Figs$.$~\ref{fig:fig2}(c),\ref{fig:fig2}(d) and~\ref{fig:fig3}(c) and \ref{fig:fig3}(d)). It is known that partons with lower momentum receive significant modification from elastic channels and it is this factor that washes away the expected difference between the rate sets at low $p_T$ in charged hadron \RAA.

        \section{Conclusion and Outlook} \label{sec:Conclusion}
        The observation of energy-loss of jets and charged hadrons from heavy-ion collisions 
        relative to a proton-proton baseline remains an important signal of the creation of QGP. 
        The dominant mode of energy-loss for energetic light-flavor quarks and gluons is gluon radiation as a result of soft scatterings with the medium. The interaction rate relies on the value 
        of the strong-coupling as well as the collision kernel encoding information about 
        momentum exchanges with the medium. 
        It was found that the energy-loss by the QCD jets calculated using either the LO, NLO, 
        or NP scattering kernels differed significantly when evaluated using the same value of \alphas. This is consistent with behavior of the radiative rates computed in Fig.~\ref{fig:fig1}. For example, the NLP curve for gluon emission from a gluon is $\approx 60\%$ above the 
        LO result while the NLO curve is nearly double the LO rate for $p\sim 10~T$. This is an important conclusion of our study: for a given value of the strong coupling, the intrinsic differences between results computed with the different scattering kernels considered here can be large by any objective measure. 
        
        However, in phenomenological analyses of relativistic heavy-ion collisions, the 
        overall value of $\alphas$ is typically treated as a fitting parameter. Scaling the 
        coupling value so that the new rates match the LO rate at large values of 
        temperature-scaled momentum ($p/T$) brings them much closer to the LO result, 
        with NLO and NP nearly collapsing on each other and stand at only $\approx 40\%$ above 
        the LO curve for $g\to g+g$ at $p\sim 10~T$. The brick test using the new 
        radiative rates, excluding elastic scatterings with the medium, further demonstrated 
        this effect with the evolving parton distribution from the NLO and NP radiative rates 
        becoming nearly indistinguishable after the employed \alphas\, is rescaled (see 
        Figs$.$~\ref{fig:fig2}(a),\ref{fig:fig2}(c) and \ref{fig:fig3}(a), \ref{fig:fig3}(c)). Including elastic 
        channels [Figs.~\ref{fig:fig2}(d) and \ref{fig:fig3}(d)] brings all three rate sets even closer to each other.
        
        To study further the effect of the new collision kernels, we perform a realistic calculation 
        of parton energy-loss using the new rates. This was done for the most centralt $0-5\%$ \PbPb\, collisions at $\sqrt{s}=2.76~\mathrm{ATeV}$ using \Martini. The simulations, 
        when run with equal and fixed values of \alphas, demonstrate the same behavior as 
        the QGP brick test cases with the largest quenching done by the NLO rates. However, 
        when we fit the value of the strong coupling, separately for three rate sets, to 
        the charged hadron \RAA\, results of ALICE, ATLAS and CMS in the same collision system, 
        the resulting curves collapse on each other, becoming indistinguishable. This is because of the similarity of the scaled rates for large values of $p/T$ as well as 
        the effect of elastic scatterings in reducing the differences between different 
        radiative rates. Moreover we observe basically no change in $p_T$ 
        dependence of the charged hadron nuclear modification factor from using NLO and NP collision 
        kernels in \Martini. 
        An additional word of caution: \Martini\, as many jet-medium simulations, implements radiation as perfectly collinear with the emitting parton. This is an appropriate treatment for the LO kernel with a numerically small coupling  (with opening angle of  $O\left(g_s\right)$), the transverse momentum dynamics will need to be scrutinized for all cases with $g_s \gtrsim 1$. 
        
        As mentioned previously, we ignored the running of the strong-coupling in this study to avoid the associated complications and difficulties in interpreting our results. In a realistic run, however, \alphas\, will have a scale dependence with the physical consequence that the plasma becomes more ``transparent'' to a jet with high transverse momentum as compared with a simulation with a fixed \alphas~\cite{Park:2021yck}. Study of the new rate sets while accounting for a running coupling is currently underway.
        
        Other means of distinguishing the behavior of the three kernels and their effect 
        on the evolving distribution of hard partons in a plasma are analyses of 
        multimessenger signals \cite{Gale:2021emg}, or with more differential measurements 
        such as those involving correlations. For the former, electromagnetic probes can offer potential access to jet-medium interactions~\cite{rouz:2022}, owing to the absence of final-state interactions with the medium. Work along those lines is in progress. 

    \acknowledgments{We are grateful to Dr. Mayank Singh for his help with the \Music\, events used in this study. This work was funded in part by the Natural Sciences and Engineering Research Council of Canada, and in part by the U.S. Department of Energy, Office of Science, Office of Nuclear Physics, under grant No. DE-FG88ER40388. S.S. is grateful for support from Le Fonds de Recherche du Qu\'ebec - Nature et technologies (FRQNT), via a Bourse d'excellence pour \'etudiants  \'etrangers (PBEEE). Computations were made on the B\'eluga and Narval computers at McGill University, managed by Calcul Qu\'ebec and Compute Canada.}
    \bibliographystyle{apsrev4-1.bst}
    \bibliography{references}
\end{document}